\documentclass[prl,twocolumn,floatfix,preprintnumbers,showpacs]{revtex4}
\usepackage{graphicx}
\usepackage{dcolumn}
\usepackage{bm}
\usepackage{color,amsmath}
\usepackage{hyperref}

\begin{document}
\preprint{LAPTH-1210/07, arXiv:0710.1630}
\title{What do WMAP and SDSS really tell about inflation?} 
\author{Julien Lesgourgues $^a$\footnote{julien.lesgourgues@lapp.in2p3.fr}, Alexei A. Starobinsky $^b$\footnote{alstar@landau.ac.ru} and Wessel Valkenburg $^a$\footnote{wessel.valkenburg@lapp.in2p3.fr}}

\affiliation{$^a$LAPTH\footnote{Laboratoire de Physique
Th\'eorique d'Annecy-le-Vieux, UMR5108}, Universit\'e de Savoie \&
CNRS, 9 chemin de Bellevue, BP110, F-74941 Annecy-le-Vieux Cedex,
France} 
\affiliation{$^b$Landau Institute for Theoretical Physics, Russian Academy
of Sciences, Moscow 119334, Russia}
\date{\today}
\begin{abstract}
We derive new constraints on the Hubble function $H(\phi)$ and
subsequently on the inflationary potential $V(\phi)$ from WMAP 3-year
data combined with the Sloan Luminous Red Galaxy survey (SDSS-LRG),
using a new methodology which appears to be more generic, conservative
and model-independent than in most of the recent literature, since it
depends neither on the slow-roll approximation for computing the
primordial spectra, nor on any extrapolation scheme for the potential
beyond the observable e-fold range, nor on additional assumptions
about initial conditions for the inflaton velocity.  This last feature
represents the main improvement of this work, and is made possible by
the reconstruction of $H(\phi)$ prior to $V(\phi)$.  Our results only
rely on the assumption that within the observable range, corresponding
to $\sim$ 10 e-folds, inflation is not interrupted and the function
$H(\phi)$ is smooth enough for being Taylor-expanded at order one, two
or three.  We conclude that the variety of potentials allowed by the
data is still large. However, it is clear that the first two slow-roll
parameters are really small while the validity of the slow-roll
expansion beyond them is not established.
\end{abstract}

\maketitle
Cosmic inflation was introduced as a simple and aesthetically elegant scenario
of the early Universe evolution which is capable of explaining its main properties
observed at the present time
\cite{Starobinsky:1980te,Guth:1980zm,Sato:1980yn,Linde:1981mu,Albrecht:1982wi,Linde:1983gd}.
As a very important byproduct it provides a successful mechanism for the
quantum-gravitational generation of primordial scalar (density) perturbations
and gravitational waves
\cite{Starobinsky:1979ty,Mukhanov:1981xt,Hawking:1982cz,Starobinsky:1982ee,Guth:1982ec,Bardeen:1983qw,Abbott:1984fp}. The Fourier power spectrum ${\cal P}_{\cal R}(k)$ of the
former ones is observed today
in the cosmic microwave background (CMB) and the large scale structure (LSS).
Vice versa, at present the CMB and the LSS provide the only quantifiable observables
which can confirm or falsify inflationary predictions. That is why matching concrete
inflationary models to observations has become one of the leading quests in
cosmology.

In the simplest class of inflationary models, inflation is driven by a
single scalar field $\phi$ (an inflaton) with some potential $V(\phi)$
which is minimally coupled to the Einstein gravity. For these models,
some new conservative bounds on $V(\phi)$ were presented recently
in~\cite{Lesgourgues:2007gp}. 
Until then, most post-WMAP3
studies concerning $V(\phi)$ relied on the slow-roll
approximation in the calculation of perturbation power spectra and
their relation to values of $\phi$ during inflation 
\cite{Spergel:2006hy,Peiris:2006ug,deVega:2006hb,Easther:2006tv,Kinney:2006qm,Finelli:2006fi,Peiris:2006sj,Cardoso:2006wf,Destri:2007pv}, 
or made
an extrapolation of $V(\phi)$ from the observable window till the
end of inflation \cite{Martin:2006rs,Ringeval:2007am,Powell:2007gu}
(a numerical integration of exact wave equations for perturbations to
obtain primordial power spectra was also permormed in 
Refs.~\cite{Makarov:2005uh,Covi:2006ci,Lorenz:2007ze} for
specific inflationary models).
The extrapolation over the full duration
of inflation is more constraining than the data alone. Instead,
Ref.~\cite{Lesgourgues:2007gp} focused only on the observable part of
the potential to see up to what extent current data really constrains
inflation.

 For this class of models, the evolution of a spatially flat Friedmann-Lemaitre-Robertson-Walker (FLRW) universe can be described by 
\cite{Muslimov:1990be,Salopek:1990jq}
\begin{align}
\dot \phi &= -\frac{m_P^2}{4 \pi} H'(\phi)\label{eq:eom}\\
 -\frac{32\pi^2}{m_P^4}V(\phi)&=\left[H'(\phi)\right]^2-\frac{12\pi}{m_P^2}H^2(\phi).
\label{eq:flrw} 
\end{align}
whenever $\dot\phi\not= 0$ and not specifically during inflation (so $
H'(\phi) \not= 0$, too). Here $H(\phi(t))\equiv \dot a/a$, $a(t)$ is
the FLRW scale factor, a dot denotes the derivative with respect to
the cosmic time $t$, a prime with respect to an
argument, and we have set $Gm_P^2=\hbar=c=1$.  If $V(\phi)$ is considered
as the defining quantity, the initial conditions for generating the
observable window are determined by the set $\{\dot \phi_{\rm ini},
V(\phi)\}$. In Ref.~\cite{Lesgourgues:2007gp}, the inflaton potential
was parametrized as a Taylor expansion up to some order, to see up to
what extent the potential can be constrained by pure
observations. However, in order to reduce the number of free
parameters, $\dot \phi_{\rm ini}$ was fixed for each model by
demanding that the inflaton follows its attractor solution just when
the observable modes exit the horizon.  In practice this means that
the results of Ref.~\cite{Lesgourgues:2007gp} assumed that inflation
started {\em at least} a few e-folds before the observable modes left
the horizon.  These precurring e-folds led to a slightly stronger
bound on the potentials than the data itself could actually give,
although this extra constraining power stands in no proportion to an
extrapolation over the full duration of inflation.

Eqs.~(\ref{eq:eom},~\ref{eq:flrw}) however show that when one
considers $H(\phi)$ as the defining quantity, all initial conditions
are already uniquely set by $H(\phi)$. Moreover, the slow-roll conditions
which require, in particular, that the first term in the rhs of
Eq.~(\ref{eq:flrw}) is much less than the last one need not be imposed
{\it ab initio}. In this Letter we derive the
bounds on $H(\phi)$ during observable inflation using its
Taylor expansion at various orders.
We infer for this some constraints on $V(\phi)$
under an even more conservative approach than in
Ref.~\cite{Lesgourgues:2007gp}, since the present method
requires absolutely no extrapolation outside
of the observable region (either forward or backward in time).  Our
only restriction is to assume that observable cosmological perturbations
originate from the quantum fluctuations of a single inflaton field,
which dynamics during observable inflation is compatible with
a smooth, featureless $H(\phi)$. 

{\em Method.} We used the publicly available code {\sc
  cosmomc}~\cite{Lewis:2002ah} to do a Monte Carlo Markov Chain (MCMC)
  simulation. We added a new module (released at 
  \url{http://wwwlapp.in2p3.fr/~valkenbu/inflationH/})
  which computes numerically the
  primordial spectrum of scalar and tensor perturbations for each
  given function $H(\phi-\phi_*)$, where $\phi_*$ is an arbitrary
  pivot scale in field space. This module is simpler than the one in
  Ref.~\cite{Lesgourgues:2007gp}, since the code never needs to find
  an attractor solution of the form $\dot{\phi}(\phi)$. The comoving
  pivot wavenumber is fixed once and for all to be
  $k_*=0.01~$Mpc$^{-1}$, roughly in the middle of the observable
  range. Primordial power spectra are computed in the range $[k_{\rm
  min}, k_{\rm max}]=[5\times 10^{-6},5]~$Mpc$^{-1}$ needed by {\sc
  camb}, imposing that $k_*$ leaves the Hubble radius when
  $\phi=\phi_*$. In practice, this just means that for each model the
  code normalizes the scale factor to the value $a_* = k_* / H_*$ when
  $\phi=\phi_*$. Note that by mapping a window of inflation to a
  window of observations today, our approach is independent of the
  mechanism of reheating. The evolution of each scalar/tensor 
  mode is given by
  \begin{align}
    \frac{d^2 \xi_{\rm S,T}}{d\eta^2} + \left[ k^2 - \frac{1}{z_{\rm S,T}}
    \frac{d^2 z_{\rm S,T}}{d\eta^2}\right] \xi_{\rm S,T}=0
  \end{align}
  with $\eta=\int dt/a(t)$ and $z_S= a\dot\phi/H$ for scalars, 
  $z_T= a$ for tensors.
  The code integrates this equation starting from the initial condition
  $\xi_{\rm S,T} = e^{-ik \eta}/\sqrt{2k}$ when $k/aH=50$,
  and stops when the expression for the
  observed scalar/tensor power spectrum freezes out in the long-wavelength regime.
  More precisely, the spectra are given by
  \begin{align}
  \frac{k^3}{2 \pi^2} \frac{|\xi_S|^2}{z_S^2} \rightarrow {\cal P}_{\cal R}~,
  \qquad
  \frac{32 k^3}{\pi m_P^2} \frac{|\xi_T|^2}{z_T^2} \rightarrow {\cal P}_{h}~,
  \end{align}
  and integration stops when $[ d \ln {\cal P}_{{\cal R}, h} / d \ln a ] < 10^{-3}$.
  If for a given
  function $H(\phi-\phi_*)$ the product $aH$ cannot grow enough for
  fullfilling the above conditions, the model is rejected. In
  addition, we impose that $aH$ grows monotonically, which is
  equivalent to saying that inflation is not interrupted during the
  observable range. If these conditions are satisfied, the power
  spectra are compared to observations.

  We choose to parametrize $H$ as a Taylor expansion with respect to
  $\phi-\phi_*$ up to a given order $n$ varying between one and three
  (this choice of background parametrization is equivalent to
  that in Ref.~\cite{Easther:2006tv}, as long as no extrapolation is
  made).  Note that for $n>1$ such an assumption excludes $\dot\phi$
  and $H'$ becoming zero at some value $\phi=\phi_1$ in the range
  involved since then $H(\phi)$ would acquire a non-analytic part
  beginning from the term $\propto |\phi-\phi_1|^{3/2}$ (with
  $V(\phi)$ being totally analytic at this point)~\footnote{The case
  of $\dot\phi$ becoming zero at the beginning or during inflation
  requires special consideration, see
  \cite{Starobinsky:1996ek,Seto:1999jc} in this respect.}. As a
  cosmological background we used the standard $\Lambda$CDM-model with
  the free parameters shown in Table~\ref{table:params}.
\begin{table}
\begin{center}
\begin{tabular}{l|ccc}
Parameter & $n=1$ & $n=2$ & $n=3$ \\
\hline
\hline
$\Omega_b h^2$ & 
$0.023 \pm 0.001$ & $0.023 \pm 0.001$ & $0.022 \pm 0.001$ \\
$\Omega_{cdm} h^2$ & 
$0.109 \pm 0.004$ & $0.109 \pm 0.004$ & $0.110 \pm 0.004$ \\
$\theta$ & 
$1.042 \pm 0.003$ & $1.041 \pm 0.004$ & $1.040 \pm 0.004$ \\
$\tau$ & 
$0.08 \pm 0.03$ & $0.08 \pm 0.03$ & $0.09 \pm 0.03$ \\
$\ln \! \left[\frac{4 H^4_*}{H'^2_* m_P^6}10^{10}\right]$ &
 $3.07 \pm 0.06$ & $3.07 \pm 0.06$ & $3.09 \pm 0.06$ \\
$\left(\frac{H'_*}{H_*}\right)^2 \!\!\! m_P^2$ & 
$0.079 \pm 0.031$ & $0.072 \pm 0.056$ & $0.081 \pm 0.067$ \\
$\frac{H''_*}{H_*} m_P^2$ & 
$0$ & $-0.035 \pm 0.199$ & $-0.079 \pm 0.247$ \\
$\frac{H'''_*}{H_*} \frac{H'_*}{H_*} m_P^4$ & 
$0$ & $0$ & $1.53 \pm 1.23$ \\
\hline
$- \ln {\cal L}_{\rm max}$ & 1781.7 & 1781.4 & 1780.1 \\
\end{tabular}
\end{center}
\caption{Bayesian 68\% confidence limits for $\Lambda$CDM inflationary
models with a Taylor expansion of $H(\phi-\phi_*)$ at order $n=1,2,3$
(with the primordial spectra computed numerically). The last line
shows the maximum likelihood. The first four
parameters have standard definitions (see e.g. \cite{Lesgourgues:2007gp}).
}\label{table:params}
\end{table}

\begin{figure}[t]
\includegraphics[width=9cm]{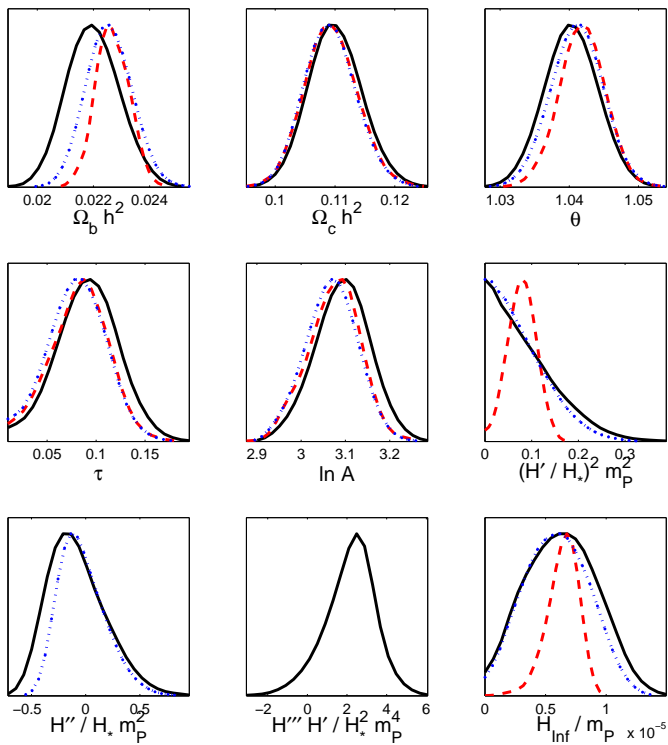}
\caption{Probability distribution for the eight independent parameters
of the models considered here,
normalized to a common arbitrary value of ${\cal P}_{\rm max}$. 
The ninth plot shows a related
parameter (with non-flat prior): namely, the value of the expansion
rate when the pivot scale leaves the horizon during inflation.  Our
three runs $n=1, 2, 3$ correspond respectively to the dashed red,
dotted blue and solid black lines.  The data consists of the WMAP
3-year
results~\cite{Spergel:2006hy,Page:2006hz,Hinshaw:2006ia,Jarosik:2006ib}
and the SDSS LRG spectrum~\cite{Tegmark:2006az}. The first four
parameters have standard definitions (see e.g. \cite{Lesgourgues:2007gp}),
and $\ln A$ is a shortcut notation
for the parameter defined in the fifth line of Table~I.
}\label{fig:1d}
\end{figure}

{\em Results for $H(\phi-\phi_*)$.} In Fig.~\ref{fig:1d} we show the
probability distribution of each parameter marginalized over the
other parameters.
The corresponding 68\% confidence limits are displayed in
Table~\ref{table:params}, as well as the minimum of the effective
$\chi^2$ for each model. This minimum does not decrease significantly
when $n$ increases, which reflects the fact that current data are
compatible with the simplest spectra and potentials, but derivatives
up to $H'''$ can be constrained with good accuracy. Note that it would
be very difficult to give bounds directly on the set $\{H, H', H'',
H''', ...\}$: indeed, these parameters are strongly correlated by the
data, because physical effects in the power spectra depend on
combinations of them.  For example, at the pivot scale, the scalar
amplitude is mainly determined by $({H^2_*}/{H'_*})^2$ and the
tensor-to-scalar ratio $r\equiv {\cal P}_h/{\cal P}_{\cal R}$ by
$({H'_*}/{H_*})^2$.  The scalar tilt
$n_S$ further depends on ${H''_*}/{H_*}$, and the scalar running on
${H'''_* H'_*}/{H^2_*}$. The Markov Chains can converge in a
reasonable amount of time only if the basis of parameters (receiving
flat priors) consists in functions of each of the above quantities, or
linear combinations of them. However, we also show in the last plot of
Fig.~\ref{fig:1d} the distribution of $H_*$: this information is
useful since the energy scale of inflation is given by $\lambda = (3
H_*^2 m_P^2 / 8 \pi)^{1/4}$, but the displayed probability should be
interpreted with care since this parameter has a non-flat prior.

The run $n=1$ is not very interesting. Indeed, imposing $H''$ and
higher derivatives to vanish leads to a one-to-one correspondence (at
least in the slow-roll limit) between the amplitude and the tilt of
the scalar spectrum. This feature is rather artificial and
unmotivated. It explains anyway why the parameter $H_*$ has
exceptionally a lower bound in the $n=1$ case~\footnote{Both $\ln A$
(the scalar amplitude) and $n_S-1$ (the scalar tilt deviation from
one) are bounded by the data. In the $n=1$ case, these two quantities
derive from $H_*$ and $H'_*$, which are hence both constrained
independently of each other.}.  Much more interesting is the $n=2$
case for which the tensor ratio, scalar amplitude and scalar tilt are
completely independent of each other, and the $n=3$ case for which
even the tilt running has complete freedom. The runs for $n=2$ and
$n=3$ nicely converged and constitute the main result of this work.
Note also that the middle-right and lower-right graphs in Fig. 1 are
compatible with each other in the following sense: though $H'*$ may
not reach zero under our assumption, the quantity $H'*/H*$ may be
arbitrarily small if $H*$ is allowed to be arbitrarily small,
too. Thus, for cases $n=2,3$ when $H*$ is not suppressed at zero
argument, $H'*/H*$ is not suppressed there, too.

The probability distribution for combinations of $H_*$, $H'_*$ and
$H''_*$ are robust in the sense that they do not change significantly
when one extra free parameter $H'''_*$ is included: this indicates
that they are directly constrained by the data. We tried to include an
additional parameter $({H''''_*}/{H_*}) ({H'_*}/{H_*})^2 m_P^6$, but
then our Markov Chains did not converge even after accumulating of the
order of $10^5$ samples. We conclude that current data do not have the
sensitivity required to constrain $H(\phi)$ beyond its third
derivative and to establish the validity of the slow-roll
approximation beginning from this order. On the other hand, the first
two slow-roll parameters $\epsilon(\phi)=H'^2m_P^2/4\pi H^2$ and
$\tilde\eta(\phi)=H''m_P^2/4\pi H$ are really small over the observed
range (tilde is used here to avoid mixing with the conformal time
$\eta$).  The next parameter $\xi \equiv
{ }^2 \! \lambda_H=H'''H'm_P^4/(4 \pi)^2 H^2$ 
is also small, $\sim 0.01$, though being
of the order of $\epsilon$ and $|\tilde \eta|$, not $\epsilon^2$ or
$\tilde \eta^2$ as would follow from the standard slow-roll
expansion. This smallness explains why our results for these
parameters are similar to those obtained for the same background
$H(\phi)$ but using the slow-roll approximation to calculate the power
spectra \cite{Peiris:2006sj} (and to those in \cite{Martin:2006rs},
too) although some important differences exist.

\begin{figure}[t]
\includegraphics[width=9cm]{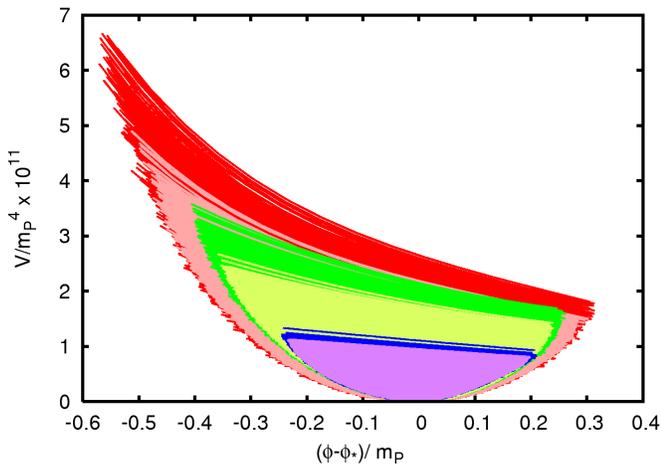}
\caption{Allowed inflationary potentials $V(\phi-\phi_*)$ inferred
from each of our $n=1,2,3$ runs. For each case, the light colour
corresponds to models allowed at the 68\% confidence levels, and the
dark colour to the 95\% level. The inner (bluish) region is obtained
for $n=1$, the intermediate (greenish) one for $n=2$ and the outer
(reddish) one for $n=3$. Each potential is plotted between the two
values $\phi_{1}$ and $\phi_{2}$ corresponding to Hubble exit for the
limits of the observable range $[k_1, k_2]$=$[2 \times 10^{-4},
0.1]~$Mpc$^{-1}$: so we
only see here the actual observable part of each potential.  Note that
this figure shows only one half of the possible solutions: the other
half is obtained by reflection around
$\phi=\phi_*$.}\label{fig:parrotstail}
\end{figure}

{\em Results for $V(\phi-\phi_*)$.}  We further processed our $n=1, 2,
3$ runs in order to reconstruct the inflaton potential. For each run,
we kept only 68\% or 95\% of the models with the best likelihood, and
computed the corresponding inflaton potentials using
Eq.~(\ref{eq:flrw}).  Note that the problem is fully symmetric under
the reflection
$(\phi-\phi_*) \leftrightarrow -(\phi-\phi_*)$. We choose to focus
on one half of the solutions, corresponding to $\dot{\phi}>0$ and 
hence $V'_*>0$.  Our
results are shown in Fig.~\ref{fig:parrotstail}.
They appear to be compatible with those of
Ref.~\cite{Lesgourgues:2007gp}, although a detailed comparison is
difficult: first, the current method is more conservative, and second,
a given order in the Taylor-expansion of $H(\phi-\phi_*)$ is not
equivalent to another order in that of $V(\phi-\phi_*)$.  Our
results are also difficult to compare with those of
Ref.~\cite{Powell:2007gu}, since these authors choose to present their
full allowed potentials extrapolated till the end of inflation: in
principle, our Fig.~\ref{fig:parrotstail} can be seen as a
zoom on the directly constrained, small $\phi$ region in their Fig.~2.

\begin{figure}[h]
\includegraphics[width=9cm]{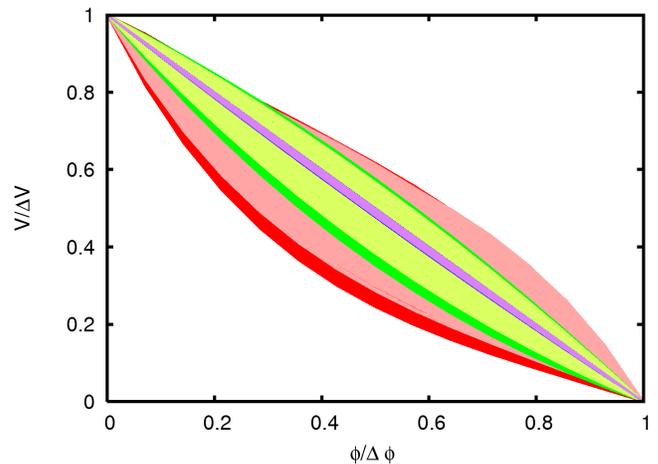}
\caption{Allowed inflationary potentials $V(\phi-\phi_*)$ with the same
colour/shade code as in Fig.~\ref{fig:parrotstail}, but a different choice
of axes: each potential is now rescaled to the same variation in $V$ an
$\phi$ space. This shows that many allowed potentials are actually convex.
The outer region still corresponds $n=3$, the intermediate one to $n=2$
and the inner (quasi-linear) one to $n=1$.
}\label{fig:feather}
\end{figure}

Our results could give the wrong impression that
all preferred potentials are concave. This comes from the fact that in
the representation of Fig.~\ref{fig:parrotstail}, many interesting
potentials are hidden, since they almost reduce to the point $(V_*,
\Delta \phi) \rightarrow (0,0)$. Indeed, as long as the
tensor-to-scalar ratio is not bounded from below, many low-energy
inflationary models with very small $H_*$ and $H'_*$ (and hence tiny
variation of the inflaton field during the observable e-folds) are
perfectly compatible with observations. It is straightforward to show
that models leading to $n_S<1$ and small $r$ correspond to convex
potentials (like e.g. new inflation with $V=V_0-\lambda \phi^n$, or
one-loop hybrid inflation with $V=V_0+ \lambda \ln \phi$), while
models with same $n_S$ and larger $r$ derive from concave potentials
(like e.g. monomial inflation $V=\lambda \phi^\alpha$). Current data
favor $n_S<1$, and the upper bound on $r$ is too loose for
differentiating between these two situations. So, our allowed
potentials can be split in two subsets: low-energy convex potentials
and high-energy concave potentials, as illustrated in
Fig.~\ref{fig:feather}, in which we rescaled all allowed potentials to
the same variation in $V$ and $\phi$.
More generally, this large degeneracy in potential
reconstruction reflects the fact that an infinitely precise
measurement of the scalar spectrum ${\cal P}_{\cal R}$ would only
constrain the function
\begin{equation}
{\cal P}_{\cal R}(k) = \left. \frac{4 H^4}{m_P^4 H'^2}
\right|_{k=aH}
\end{equation}
(in the slow-roll approximation).
This is not sufficient for inferring the correspondence between $k$ and
$\phi$, and hence for
a unique determination of $H(\phi)$ and $V(\phi)$. It is
necessary to measure also the tensor spectrum, equal to
\begin{equation}
 {\cal P}_{h}(k) = \left. \frac{16 H^2}{\pi m_P^2}\right|_{k=aH}
\end{equation}
in the same approximation, in order to diminish this degeneracy
(see the related discussion in Ref.~\cite{Cline:2006db}). In the
slow-roll approximation, the knowledge of ${\cal P}_{h}(k)$ leads to
the unambiguous determination of $H(\phi)$. However, the question how
unique the determination of $H(\phi)$ is, even from {\em both}
${\cal P}_{\cal R}(k)$ and ${\cal P}_{h}(k)$ in the generic case beyond
slow-roll, is still open because of the existence of many $H(\phi)$ leading
to the same perturbation spectra which may not be obtained from the
slow-roll expansion at all ~\cite{Starobinsky:2005ab}. Still, since the
difference of these additional solutions from slow-roll ones
is, in some sense, exponentially small for small slow-roll parameters,
their existence might appear not significant from the observational point
of view.


\section*{Acknowledgements.}
This work follows from a very nice and fruitful stay of JL and WV 
at the Galileo
Galilei Institute for Theoretical Physics, supported by INFN. JL and
WV also wish to thank Prof. Lev Kofman for very useful discussions.
AS was partially supported by the Russian Foundation for Fundamental Research,
grant 05-02-17450, and by the Research programme``Elementary particles'' of
the Russian Academy of Sciences. He also wishes to thank Prof. Alikram Aliev and
the Feza G\"ursey Institute, Istanbul, for hospitality during the period of
completion of this paper. WV is
supported by the EU 6th Framework Marie Curie Research and Training
network ``UniverseNet'' (MRTN-CT-2006-035863). Numerical simulations
were performed on the MUST cluster at LAPP (CNRS \& Universit\'e de
Savoie).

\bibliography{refs}
\end{document}